# Dark plasmons in hot spot generation and polarization in interelectrode nanoscale junctions

*Joseph B. Herzog,*[†,#] *Mark W. Knight,*[‡,‖,#] *Yajing Li,*[†] *Kenneth M. Evans,*[†] *Naomi J. Halas,*[†,‡,‖,§] *Douglas Natelson\**[†,‡]

[†]Department of Physics and Astronomy, [‡]Department of Electrical and Computer Engineering, [‖]Laboratory for Nanophotonics, [§]Department of Chemistry, Rice University, 6100 Main Street, Houston, Texas, 77005, United States

ABSTRACT: Nanoscale gaps between adjacent metallic nanostructures give rise to extraordinarily large field enhancements, known as "hot spots", upon illumination. Incident light with the electric field polarized *across* the gap (along the interparticle axis) is generally known to induce the strongest surface enhanced Raman spectroscopy (SERS) enhancements. However, here we show that for a nanogap located within a nanowire linking extended Au electrodes, the greatest enhancement and resulting SERS emission occurs when the electric field of the incident light is polarized *along* the gap (transverse to the interelectrode axis). This surprising and counterintuitive polarization dependence results from a strong dipolar plasmon mode that resonates transversely across the nanowire, coupling with dark multipolar modes arising from subtle intrinsic asymmetries in the nanogap. These modes give rise to highly reproducible SERS enhancements at least an order of magnitude larger than the longitudinal modes in these structures.

Plasmons in metal nanostructures are crucial for understanding the physics of nanoscale optics. These charge density fluctuations on metal surfaces couple with incident light and generate local optical fields that are strongly enhanced relative to those of the incident radiation. Plasmonic enhancements in nanostructures have enabled surface enhanced Raman spectroscopy (SERS)[1,2] along with promising applications including single molecule studies,[3,4] plasmonically enhanced photovoltaics,[5–7] deep subwavelength lasing,[8–10] plasmon-enhanced photodetection,[11–13] and ultra-high-resolution color printing.[14]

SERS in nanogaps has been studied extensively, including investigations on the polarization dependence of the enhanced Raman excitation and emission. SERS emission from the nanojunction between finite, subwavelength dimer structures consistently shows its strongest magnitude when the incident light is polarized across the gap (along the interparticle axis).[15–20] This is also true for gaps in bowtie nanoantennas[21] and gaps between a single nanoparticle on the side edge of a larger nanowire.[22,23]

Polarization studies have also been performed on plasmonically active structures with extended electrodes, which are critical for hybrid optical/electronic applications of these systems. Electromigrated nanojunctions with wide (~20 nm) gaps have shown strongest enhancements with light polarized across the gap.[24] SERS signals in mechanical break junctions have also been shown to be strongest when incident light is polarized across the gap.[25,26] Theoretical models have confirmed these observations, which are accurate for perfectly smooth nanogaps with an idealized geometry.[27] Other extended electrode studies, including tip-enhanced measurements, typically use this same across-the-gap polarization, assuming that the strongest enhancement occurs with this polarization direction,[28–30] in part due to the "lightning rod effect."[31]

Here we show that both electromigrated and self-aligned nanojunctions of particular dimensions give rise to extremely large SERS signals when the incident light is polarized with the electric field *along* the nanogap (*perpendicular* to the interelectrode axis). In our structures light with this transverse polarization excites complex hybridized plasmon modes with greatly enhanced local fields, allowing for much stronger SERS emission than seen in the same structures using across-the-gap excitation polarization. Cathodoluminescence imaging of the bright modes of the electrode junction, combined with finite element calculations, allows us to identify the hybridized plasmon modes of the self-aligned electrode structure responsible for the unusual and counterintuitive property of these structures.

The nanojunctions studied in this work were created with two different fabrication methods: electromigration[32] and a self-aligned approach.[33] Self-aligned nanogaps have several advantages over electromigrated junctions, including the capability of mass-producing plasmonically active SERS hotspots with large sensitivities similar to those seen in electromigrated gaps. An entire array of self-aligned junctions can be fabricated in parallel with only two lithography and evaporation steps. In contrast, the fabrication scalability of electromigrated junctions is limited, since each individual junction needs to be electromigrated separately. The self-aligned technique also allows for greater systematic nanoscale control of the gap geometry, avoiding the randomness inherent in the electromigration process that often creates asymmetrical gaps. Finally, self-aligned devices are more robust than electromigrated gaps, having shelf lives that can exceed a year at ambient conditions, while electromigrated junctions typically lose their enhanced Raman response in a day due to structural relaxation of the metal configuration. Despite these differences, both the electromigration[28,34–36] and self-aligned[37] techniques successfully produce nanogaps which demonstrate strong SERS enhancement, with similar microscale geometries and the same metallization.

The electromigrated devices are fabricated with the same technique described in detail in previous work.[34–36,38] A summary of the procedure is as follows. A single electron-beam lithography (EBL) step patterns the device on a doped silicon substrate with a 200 nm thermal silicon oxide. E-beam evaporation of a 1 nm Ti adhesion layer and 15 nm of Au, followed by with liftoff processing, creates a 120 nm wide by 700 nm long Au nanowire connecting two large triangular electrodes. After oxygen plasma cleaning, the device is ready for self-assembly of molecules. In this study we soak the samples in a 0.1 mM solution of trans-1,2-bis(4-pyridyl)-ethylene (BPE) in ethanol for 45 min, which leaves a self-assembled monolayer of BPE on the surface. Finally, the nanowires are electromigrated one at a time to form the nanojunctions with gaps ranging from 2-10 nm.

Self-aligned junctions are fabricated with a two-step lithography process, initially developed in previous work.[33] The first lithography step patterns the left side of the nanowire and left electrode on a doped silicon substrate with a 200 nm thermal silicon oxide. After developing, four layers are evaporated: 1 nm Ti, 15 nm Au, 1 nm $SiO_2$, and 12 nm of Cr. Ti is used as an adhesion layer, and Au is the plasmonically active metal used for the device. The $SiO_2$ acts as a barrier to prevent the Cr from diffusing into the Au and altering the gold's optical properties,[39] and the Cr layer is crucial for the self-aligning process. After evaporation, the chromium layer oxidizes and swells, creating a chromium-oxide ledge extending a few nanometers beyond the metal layers. The Cr-oxide overhang acts as a shadow mask for the subsequent evaporation. After liftoff, the second EBL step patterns the right side of the nanowire overlapping the first side, as well as the other electrode. A subsequent evaporation deposits the same four layers, and the overlapping pattern along with Cr-oxide-mask "self-align" the two sides, creating the nanogaps. A Cr etch follows liftoff, which removes the overlapping material along with all the Cr. Finally, the $SiO_2$ barrier layer is etched away with a brief buffered oxide etch, leaving a clean a 2-10 nm gap at the center of a 700 nm long and 120 nm wide nanowire connected to two Au electrodes. The devices are now finished and ready for molecules to be self-assembled on the surface. The optical microscope image (Figure 1a) shows an overview of a typical self-aligned structure, and the scanning electron microscope image (Figure 1b) displays the nanogap at the center of the nanowire.

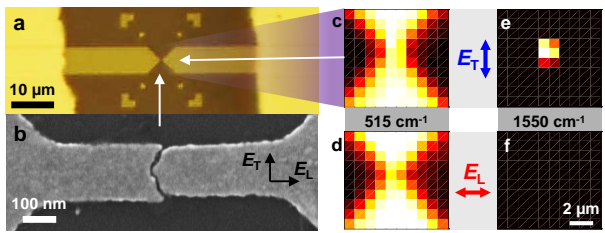

**Figure 1.** (a) Optical microscope image showing overview of device with nanogap at the center of the Au bowtie-electrodes. (b) Scanning electron microscope (SEM) image of typical self-aligned Au nanowire with nanogap, and a legend that shows the definitions of the transverse, $E_T$, and longitudinal, $E_L$, polarization orientations. (c-f) Raman spectral maps at center of bowtie where pixel-brightness is proportional to Raman intensity. All maps are of the same nanogap device and image the same area with scale-bar as indicated in (f). Spectral maps of the Raman mode of Si (510-520 $cm^{-1}$) for (c) transverse and (d) longitudinal polarizations plotted with the same intensity color scale. Spectral maps of SERS of BPE, integrated from 1500-1600 $cm^{-1}$, for

(e) transverse and (f) longitudinal polarizations, both plotted with the same intensity-color scale. Map (e) shows SERS hotspot at gap is enhanced with transverse polarization, while map (f) is dark, indicating that the hotspot is not detectable with longitudinal polarization.

Optical measurements were made with a custom Raman microscope, as described in previous work.[34,36] A 785 nm laser is rastered across the sample with submicron resolution and focused on the sample with a 100x (N.A. = 0.7) long working distance objective. By taking a spectrum at each position of a spatial grid, spectral maps can be created. Figure 1c-f has four of these maps of the same device and area, where each intensity pixel was obtained from a single spectrum with a 1 s integration time. All maps plot the integrated Raman intensity for a specific spectral range. Figures 1c-d plot the integrated spectra from 510 to 520 $cm^{-1}$, the Raman response of the Si substrate; and Figures 1e-f map the SERS signal from BPE in the range 1500-1600 $cm^{-1}$. The polarization of the illuminating laser light was rotated with a half-wave plate in order to create spatial maps as a function of incident polarization. Maps in Figure 1c,e were each obtained from data from the same scan using transversely polarized incident light, and spectral maps in Figure 1d,f were from data of a single scan obtained using longitudinally polarized incident light. The transverse, $E_T$, and longitudinal, $E_L$, polarization directions are shown relative to the self-aligned gap orientation in Figure 1b. The Si maps in Figure 1c-d show no polarization dependence; both maps are nearly identical, showing dark regions where the Au attenuates the Si Raman signal. However, the SERS hotspot at the center of the device, visible in Figure 1e, shows a strong polarization dependence, with the greatest signal being produced when the incident light is polarized transversely. With longitudinal polarization, $E_L$, the SERS signal was greatly reduced in comparison, and in some cases, as shown in Figure 1f, the hotspot signal was undetectable (all dark).

The Raman enhancement only occurs in the presence of a gap. Nanowire constrictions measured before electromigration have never shown hotpot regions. The hotpots only appear after electromigration or on the self-aligned structures fabricated with the gap, confirming that the enhancement is due to localized plasmons at the junction.

Two typical Raman spectra taken at the center of a self-aligned nanogap are plotted in Figure 2a with orthogonal incident laser polarizations, $E_T$ ($\theta = 90°$) and $E_L$ ($\theta = 0°$). The scanning electron microscope (SEM) image in Figure 2c, shows the nanogap with a polarization direction reference angle. The two spectra in (a) are the slices from the polar-contour plot of (c) marked with corresponding arrows. Each slice in the polar-contour plots, Figure 2b-d, is a single spectrum with a 3 s integration time taken at every 5 degrees of polarization angle. The shading of each plot is proportional to the Raman spectrum intensity, with brighter colors representing higher CCD counts as indicated in the color bar of each figure.

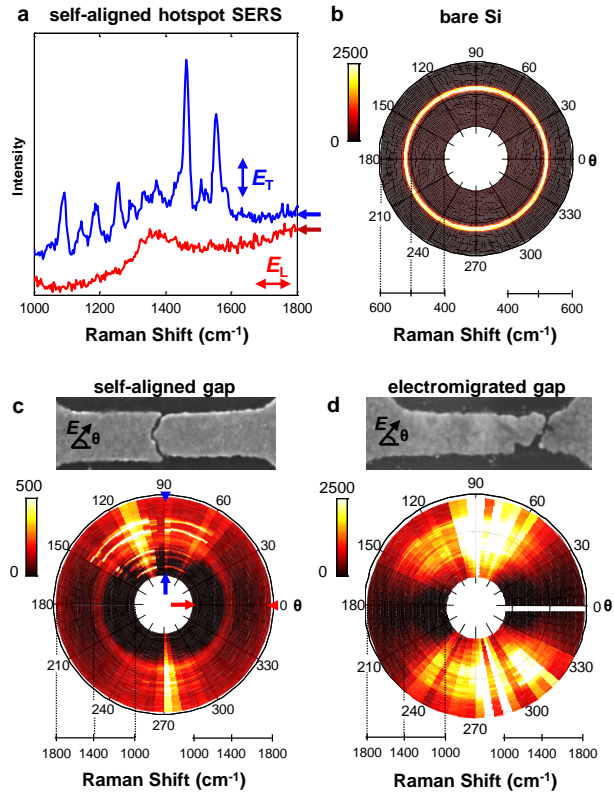

**Figure 2.** Surface enhanced Raman spectra (SERS) as a function of polarization. (a) SERS spectra of BPE at center of self-aligned nanogap with transverse, $E_T$, and longitudinal, $E_L$, incident polarization. The two spectra in (a) are the radial slices marked with arrows in the polar-contour plot (c). (b-d) Polar contour plots showing polarization dependent Raman spectra. Color shading represents Raman spectral intensity in CCD counts with scale shown in color-bar. Each radial slice represents a single spectrum taken with a 3 s integration time at every 5 degree polarization angle. All spectra are plotted in the Raman shift range that increases from inner to outer diameter as labeled on the bottom of each plot. Spectra taken away from the hotspot on nearby Si (b) show no polarization dependence. Measurements taken at (c) self-aligned and (d) electromigrated nanogap hotspots show strong polarization dependence. SEM images show a typical (c) self-aligned and (d) electromigrated gap.

Polarization-dependent Raman spectra were first obtained on a large gold pad and on bare Si near a device, which confirmed that the Raman detection system itself is polarization independent for the complete spectral range studied. The measurements for the Si peak at 515 cm$^{-1}$ are plotted in Figure 2b in the range of 400 to 600 cm$^{-1}$. These results confirm that any observed polarization dependence of the SERS signal is solely due to the nanogap and not caused by any asymmetry in the optical path of the Raman signal.

The nanogap SERS signals observed from these structures were highly polarization dependent. The polarization dependent spectra in Figure 2c-d were taken at the SERS hotspot of (c) a self-aligned and (d) an electromigrated nanojunction and are plotted in the range from 1000 to 1800 cm$^{-1}$. SEM images of each gap are shown above the corresponding polar contour plots. Both types of nanogaps showed a strong polarization dependence of the SERS signal, with the greatest intensity being measured when the incident light had a transverse (*along* the gap) polarization, $\theta = 90°$ and $\theta = 270°$. (This is in stark contrast to the many examples discussed in the introduction, which show maximum Raman response and/or field enhancement for incident light polarized *across* the gap.) Intensity fluctuations as a function of time between consecutive polarization angles are superimposed onto the SERS polarization dependence contour plots which sometimes saturated the intensity range or made the SERS signal undetectable. Since the spectra at each angle were obtained at successive times, the measured intensity as a function of angle included a fluctuating contribution due to "blinking", a typical characteristic of high sensitivity SERS signals.[34–36,38,40–42] The broad bands in the spectra of the electromigrated sample (d) are the SERS response of carbon contamination[20] deposited on the electromigrated junctions during SEM imaging prior to measurement with the Raman microscope. The carbon contamination SERS emission showed a similar polarization dependence as the BPE molecule signal in (c). Other types of molecules have also been tested in these devices with similar results.

Transversely polarized (along the gap) excitation (with $\theta = 90\pm3^{\circ}$) generated greater SERS enhancement than longitudinally polarized (across the gap) excitation on 65 of the 76 active devices measured (86%). Variances from this trend could have originated from blinking events during spectra acquisition or could be caused by defects in the metal geometry that occurred during the fabrication process, creating off-axis or opposite polarization dependence. The emitted Raman light has always been found to be partially polarized *along* the gap, transverse to the interelectrode axis ($\theta_{emitted} = 90\pm10^{\circ}$).

To further understand the origin of the observed polarization dependence and the nature of the plasmon resonances in these nanojunctions, cathodoluminescence (CL) imaging was used. Cathodoluminescence is a useful method to examine properties of the radiative plasmon modes in various nanostructures.[43–47] CL employs an SEM equipped with specialized optics to detect radiated photons in addition to ejected secondary electrons, allowing for simultaneous SEM and CL imaging. When the high-energy electron beam (30 keV) scans across a sample, it excites many plasmons modes at each beam position. The emitted photons are collected with a 0.89 N.A. parabolic mirror and reflected toward an optical spectrometer and detector system (Gatan, MonoCL4 Elite). The nanometer-sized beam focus and positioning allows for high resolution (~20 nm) imaging of the structure. CL images are excitability maps where the brightness of each pixel is proportional to the light intensity emitted from plasmon modes excited when the beam is at that particular position.

A CL excitability map and corresponding SEM image taken simultaneously are shown in Figure 3a,b. The device scanned is a typical self-aligned structure and shows a characteristic CL response typical of these nanojunctions. The CL measurement shown is unpolarized and captured with a 700 nm bandpass filter (80 nm FWHM) in order to preferentially detect plasmon emission over non-plasmonic luminescence. Though non-filtered panchromatic CL measurements have shown similar results. The CL image shows that radiative transverse plasmon modes of the nanowire dominate the signal, and that these modes are dark at the gap in the center of the nanowire. No longitudinal plasmon modes are visible. Computational modeling (see below) shows that longitudinal plasmons for this structure are comparatively weak and correspond to longer wavelengths due to the large extended electrodes on each end of the nanowire. Consistent with the measured dependence of Raman emission on the incident polarization, the strong resonant transverse plasmons dominate the excitation dependence of the SERS. The low CL signal at the gaps is due to an optically dark multipole plasmon mode which can be seen in the calculated charge plots (Figure e-iii).

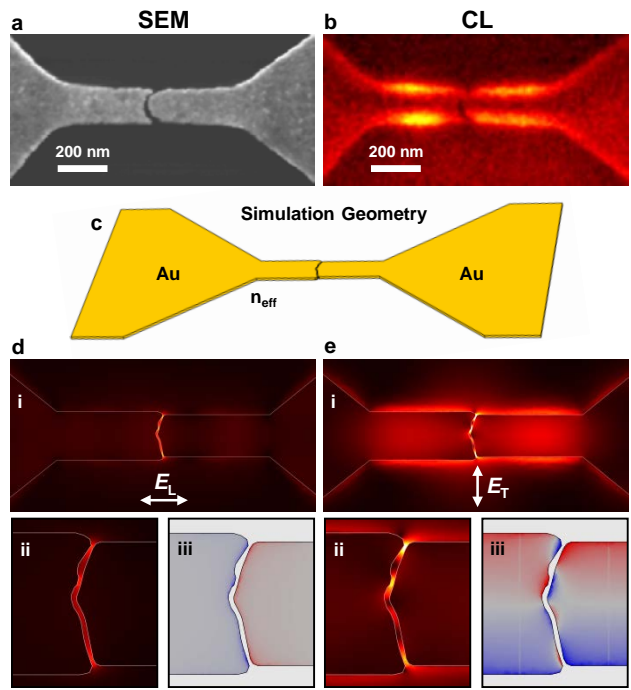

**Figure 3.** Cathodoluminescence measurements and simulations. (a) SEM image of a nanogap acquired in parallel with (b) an unpolarized cathodoluminescence excitability image. (c) Simulation geometry of self-aligned nanogap used in COMSOL finite element method calculations. Calculated electromagnetic-field enhancement (i-ii) and charge distribution (iii) due to incident Gaussian light with (d) longitudinal and (e) transverse polarization. Local, very intense multipolar plasmon modes in the gap are excited efficiently when the light is transverse polarized (*along* the gap), because that light couples well into transverse plasmon modes of the wire. The traditional longitudinal polarization (*across* the average gap direction) does not efficiently excite large local fields in the gap.

The local electromagnetic response of these structures was modeled using the finite element method (FEM, COMSOL 3.5a). The geometry of the simulated structure was designed to match, as closely as possible, a representative set of experimental dimensions. The nanowire section measured 700 nm, which expanded to a set of pads measuring 1.2 µm across. The asymmetric gap varied from 5-10 nm in width, and included several local minima in the interelectrode distance. To include the effect of the substrate, we employed an effective medium approximation with $n_{eff} = 1.25$. The incident wave was modeled as a Gaussian beam focused to the free space diffraction limit at 785 nm ($d = \lambda/2\mathrm{NA} = 560$ nm), based on the microscope objective used for the experimental measurements. The Au layer was 15 nm thick, and modeled using an experimentally derived dielectric function.[48]

The simulated response exhibits a strong contrast between longitudinal and transverse polarized excitation (Figure 3d-e). Electromagnetic-field distributions are plotted in Figures 3i-ii for (d) longitudinal and (e) transverse polarization. The electromagnetic field and charge distribution is weak for longitudinal polarization (Figure 3d). With the traditional, across-the-gap longitudinal excitation polarization, plasmons launch from the gap and then decay in the large extended electrodes, producing little *E*-field enhancement at the gap. For along-the-gap transverse polarization, the incident light excites a dipole plasmon that resonates between the edges of the nanowire, as seen in the CL image. This dipole mode interacts with subtle nanoscale structural variations in the gap, creating a highly localized hybridized mode, shown enlarged in Figure 3(e)ii-iii. When comparing these results to CL, the multipolar component of this mode is dark, consistent with the conceptual picture that the hybridized mode responsible for field enhancement is due to a coupling between non-dipole-active dark modes and the dipole active n = 1 mode. The simulation was calculated for a self-aligned structure, but any arbitrary asymmetry between the electrodes would create a similar hybridized mode; therefore, the same hybridized mode also exists in electromigrated junctions.

The polarization dependence of these devices is independent of the precise gap geometry or orientation, but rather is highly dependent on how efficiently incident light couples to the transverse dipole plasmon of the larger nanowire structure. It is that transverse plasmon that hybridizes with multipolar gap modes with large field enhancements. With our fabrication techniques gaps are formed perpendicular to the nanowire. While the gaps have some geometric variations, on average each gap lies perpendicular to the nanowire. Different devices with SEM images showing various gap geometries all have similar polarization dependence, with the strongest enhancement when the light is polarized transversely (*along* the gap).

The simulation was repeated at each polarization orientation from $\theta = 0°$ to $360°$. At each polarization angle the magnitude of the local intensity as a function of polarization angle, $G_{loc}(\theta) = E_{loc}^2(\theta)/E_0^2$, at the gap was calculated, where $E_0$ is the magnitude of the incident field. These results are shown in Figure 4b. For these structures, the local enhancement at the gap for transverse excitation has been calculated to be more than an order of magnitude larger than the enhancement with longitudinal excitation.

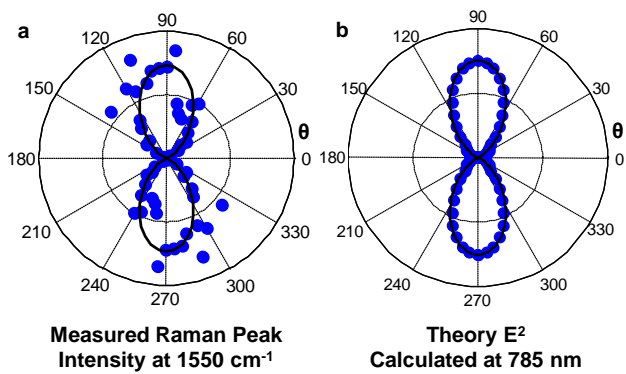

**Figure 4.** (a) Measured SERS intensity at peak position of 1550 cm$^{-1}$ as a function of incident laser polarization. (b) Calculated local enhancement of excited mode as a function of incident laser ($\lambda$ = 785 nm) polarization. Plotted is the normalized electric-field squared at the gap calculated with finite element method for simulated nanogap geometry.

The measured experimental SERS intensity as a function of incident light polarization is normalized and plotted on a traditional polar plot (Figure 4a). Each data point is the peak intensity at 1550 cm$^{-1}$ from the average of two Raman spectra with the same polarization, obtained from the raw data in the polar-contour plot of Figure 2c. Noise in the measured Raman data is largely due to spectral intensity fluctuations in time (blinking) as discussed previously. The measured polarization dependence of the SERS signal fits well with, and is directly proportional to, the calculated local enhancement. This corresponds with previous studies[23] stating that for any polarization dependent SERS measurement, the polarization dependence of the overall SERS enhancement factor, $G_{tot}(\theta)$, is directly proportional to the polarization dependence of the local intensity,

$$G_{tot}(\theta) = G_R \frac{E_{loc}^2(\theta)}{E_0^2}.$$

Here $G_R = |E_R/E_0|^2$ is the Raman emission enhancement factor, and it is polarization independent since it depends on the direction of the induced Raman dipole field, $E_R$, which is determined by the geometry of the metal structure.

Experiments on SERS in mechanical break junctions have shown a longitudinal polarization dependence, rotated 90 degrees compared to our structures, measuring strongest SERS enhancement when the light is polarized *across* the gap.[25] In their case, however, there is no nanowire connecting the wide (3-5 microns) electrodes; therefore there would be no resonant transverse dipole mode at the excitation wavelength (632 nm in their case). Moreover, in the mechanical break junction experiments the gap structure is relatively smooth and symmetrical, lacking the nanoscale asymmetries that enable comparatively large hybridization of the multipolar interelectrode plasmon with the transverse dipolar mode in our structures. Similarly, electromigrated junctions formed in chemically synthesized silver nanowires consistently show maximum Raman enhancements with the incident 514 nm light is polarized *across* the gap,[24] in contrast to our results. Again, the most likely explanation for this difference is that for the nanowire dimensions and incident wavelength, there is not a strong transverse plasmon resonance in the silver nanowires leading to the gap.

Our results emphasize that having realistic, non-idealized simulation geometries is critical for modeling the true nature of polarization dependence of SERS emission in complex structures. Often calculations assume perfect symmetry and smoothness when representing nanogaps,[25–27] while we have shown for our structures that the asymmetries of the junction and the subtle variations of gap dimensions (even when average gap orientation is preserved) are essential factors that enable the strong hybridized modes and significant polarization dependence seen in our devices.

In summary, we have shown that dark modes can play a critical role in the plasmonic hotspots in nanogap structures with extended electrodes. Nanogap hotspots formed in particular nanowire structures with extended electrodes are excited most efficiently by light polarized *along* the gap (transverse to the average interelectrode axis). This surprising behavior, which has orthogonal polarization dependence compared to all previous reported gap-enhanced Raman spectroscopy, is due to an intense transverse dipolar plasmon mode that interacts with optically dark multipolar modes localized to the nanogap region. The Raman enhancement that results from these highly localized, intense hybrid plasmon modes is stronger than that obtained for light polarized in the traditional *across*-the-gap direction. The hybridized mode has a large density of states for far-field photon emission, due to its transverse dipolar component.  Thus, while the incident polarization determines the efficiency with which a hybridized mode is excited, the Raman emission is dominated by this transverse dipolar mode, which dictates its polarization dependence. This insight into its underlying plasmonic properties explains why the nanogap structures reported here have such a robust near-infrared plasmon response: its strength is largely determined by the transverse dipolar mode of the nanowire geometry, which is highly reproducible, and not by the specific geometry of the nanogap junction. Understanding the nature of plasmons in this type of system opens the possibility of further geometric optimization of near field-far field coupling in electrode-nanogap structures, for improved Raman measurements and other nanophotonics applications.

AUTHOR INFORMATION

**Corresponding Author**

*E-mail: natelson@rice.edu

**Author Contributions**

#These authors contributed equally to this work.

**Notes**

The authors declare no competing financial interest.

ACKNOWLEDGMENT

Authors would like to acknowledge Peter Nordlander and Daniel Ward for useful conversations. JBH acknowledges supported by a Robert A. Welch Foundation postdoctoral fellowship and Lockheed Martin Corporation through LANCER. DN and YL acknowledge financial support from the Robert A. Welch Foundation (grant C-1636) and LANCER. KME acknowledges financial support from the II-VI Foundation. MWK and NJH acknowledge the Robert A. Welch Foundation under Grant C-1220, the National Security Science and Engineering Faculty Fellowship (NSSEFF) N00244-09-1-0067, the Air Force Office of Scientific Research (AFOSR) FA9550-10-1-0469, National Science Foundation Major Research Instrumentation (MRI) grant ECCS-1040478, the Defense Threat Reduction Agency (DTRA) HDTRA1-11-1-0040, the Office of Naval Research (N00014-10-1-0989), and the Army Research Office (MURI).

TOC Figure:

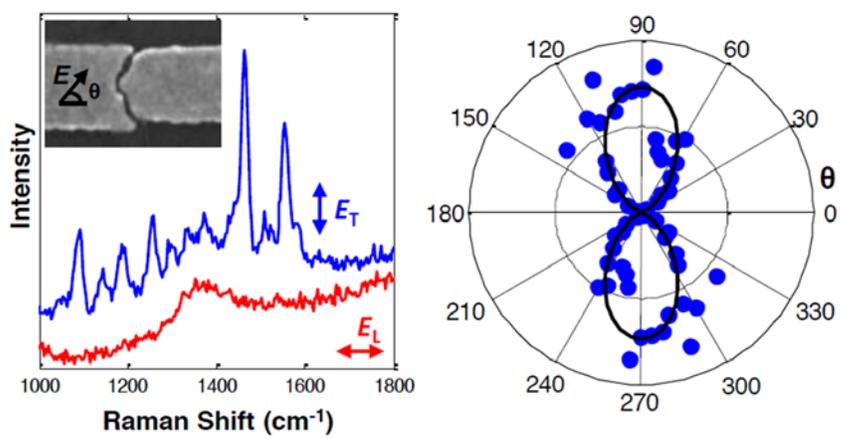